# STATISTICAL PROPERTIES OF SUPER-HOT SOLAR FLARES

AMIR CASPI[1†], SÄM KRUCKER[2,4], AND R. P. LIN[2,3,5]

[1]Laboratory for Atmospheric and Space Physics, University of Colorado, Boulder, CO 80303, USA
[†]formerly at [2]Space Sciences Laboratory and [3]Department of Physics, University of California, Berkeley, CA 94720, USA
[4]Institute of 4D Technologies, School of Engineering, University of Applied Sciences and Arts Northwestern Switzerland, 5210 Windisch, Switzerland
[5]School of Space Research, Kyung Hee University, Republic of Korea

## ABSTRACT

We use RHESSI high-resolution imaging and spectroscopy observations from ~6 to 100 keV to determine the statistical relationships between measured parameters (temperature, emission measure, etc.) of hot, thermal plasma in 37 intense (GOES M- and X-class) solar flares. The RHESSI data, most sensitive to the hottest flare plasmas, reveal a strong correlation between the maximum achieved temperature and the flare GOES class, such that "super-hot" temperatures >30 MK are achieved almost exclusively by X-class events; the observed correlation differs significantly from that of GOES-derived temperatures, and from previous studies. A nearly-ubiquitous association with high emission measures, electron densities, and instantaneous thermal energies suggests that super-hot plasmas are physically distinct from cooler, ~10–20 MK GOES plasmas, and that they require substantially greater energy input during the flare. High thermal energy densities suggest that super-hot flares require strong coronal magnetic fields, exceeding ~100 G, and that both the plasma $\beta$ and volume filling factor $f$ cannot be much less than unity in the super-hot region.

## 1. INTRODUCTION

Solar flares are characterized by the explosive release of large amounts of magnetic energy, much of which ultimately manifests as transient heating of coronal plasma to temperatures up to tens of mega-kelvin (MK), much hotter than the ambient ~1 MK temperature of the quiescent corona. Numerous observations of the X-ray signatures of hot plasma – continuum emission from free thermal electrons (bremsstrahlung, radiative recombination) and discrete line emission from bound electrons of high-charge-state thermal ions – by both broadband and Bragg crystal spectrometers (e.g., onboard *Geostationary Operational Environmental Satellite* (GOES), *Solar Maximum Mission*, *Yohkoh*, etc.) have shown that hot, ~5–25 MK thermal plasmas are ubiquitous in flares of all scales. The peak soft X-ray (SXR) flux observed by the X-ray Sensor (XRS) onboard the GOES series has become the standard measure of solar flare intensity ("GOES class"), and GOES SXR measurements are often used to derive a "bulk" temperature for the hot plasma (e.g., Garcia & McIntosh 1992; White *et al.* 2005).

The first high-resolution (~2 keV FWHM) hard X-ray (HXR; ≳20 keV) flare spectra, obtained by Lin *et al.* (1981) using cryogenically-cooled germanium detectors (GeDs), revealed for the first time a thermal component with temperatures of up to ~34 MK, significantly hotter than measurements from earlier instruments. Their precise spectra, in contrast to the coarse (Δ*E/E* of ~25% to ~133%) observations of previous broadband HXR spectrometers, allowed an accurate characterization of the steeply falling (*e*-folding of ~2 keV) HXR thermal continuum. Subsequent Fe XXVI line observations (e.g., Tanaka 1987; Pike *et al.* 1996) and inferences from broadband observations (e.g., Hudson *et al.* 1985; Lin *et al.* 1985; Jakimiec *et al.* 1988) suggested that such "super-hot" ($T_e \gtrsim 30$ MK) temperatures are common in GOES X-class flares, but no single instrument had sufficient simultaneous spectral and spatial resolution to fully characterize these thermal plasmas across a large number of events. The global properties of super-hot flares thus remain poorly known.

A number of studies have examined how the temperature of the X-ray-emitting electron population varies with the volume emission measure ($n^2V$) and GOES class (see, e.g., Garcia & McIntosh 1992; Feldman *et al.* 1996; Battaglia *et al.* 2005; Hannah *et al.* 2008; Ryan *et al.* 2012). They showed that the maximum thermal (Maxwellian) electron temperature achieved during a flare is correlated with both GOES class and emission measure. Feldman *et al.* (1996), in particular, concluded that "large" (intense) flares thus are unlikely to simply be a sum of many small, unresolved events, but rather that the flare properties must scale intrinsically. However, their analysis included temperatures only up to ~25 MK, as inferred from *Yohkoh* Bragg Crystal Spectrometer (BCS) observations, and instrumental saturation limited accuracy for the most intense flares. More recently, Ryan *et al.* (2012) used GOES self-consistently to correlate XRS-derived temperatures with GOES class and emission measure, but their maximum temperatures were only ~30 MK and also suffered from saturation for intense flares. It therefore remains an open question whether the same, or any, scaling laws extend to super-hot temperatures, and/or whether super-hot temperatures are associated with specific classes of flares.

The *Reuven Ramaty High Energy Solar Spectroscopic Imager* (RHESSI; Lin *et al.* 2002) provides high spectral and spatial resolution X-ray observations down to ~3 keV, enabling precise measurements of the thermal continuum from plasmas with temperatures ≳10 MK; it is most sensitive to the hottest plasmas, and thus is ideal for studying super-hot flares. While the ubiquitous hot, ~5–25 MK plasma is commonly accepted to result from evaporation of chromospheric material heated by collisions of flare-accelerated electrons during the impulsive phase (see the review by Fletcher *et al.* 2011), there is strong evidence that, at least during the earliest parts of the flare, the super-hot plasma is heated directly in the corona – potentially within the acceleration region – via a fundamentally different physical process (e.g., Masuda 1994; Masuda *et al.* 1998; Caspi & Lin 2010; Longcope & Guidoni 2011). It is thus unclear whether the super-hot plasma properties *should* follow the same scaling laws as non-super-hot plasma, or whether super-hot flares would constitute a unique class of event.

Here, we use RHESSI imaging and spectroscopy to survey 37 intense flares (25 M-class, 12 X-class) to obtain the maximum continuum temperature and corresponding (cotemporal) emission measure for each flare, and to compare these values to both the GOES class and the derived electron densities and thermal ener-gies. We show that the maximum flare temperature is well-correlated with GOES class, such that super-hot temperatures are associated almost exclusively with X-class flares; this correlation differs significantly from that found for GOES-derived temperatures and from those given by earlier works using GOES-, BCS-, and even RHESSI-derived temperatures. We also show that super-hot flares have ubiquitously high emission measures, electron densities, and thermal energies and energy densities – both at the time of the maximum temperature and later, when the energies are maximized – compared to non-super-hot flares. These results consistently suggest that super-hot and GOES-temperature plasmas are fundamentally dissimilar, likely resulting from different physical processes, with super-hot plasmas requiring a substantially higher total energy input. Additionally, the large thermal electron number and energy densities suggest that the plasma $\beta$ and filling factor $f$ must both be $\gtrsim 0.01$, perhaps near unity, in the region of the super-hot plasma.

## 2. OBSERVATIONAL DETAILS

The front segments of RHESSI's GeDs provide ~1 keV FWHM spectral resolution (Smith *et al*. 2002), capable of resolving and accurately measuring the steeply-falling super-hot continuum, while RHESSI's imaging spectroscopy allows characterization of both thermal and non-thermal sources with angular resolution down to ~2″ (Hurford *et al*. 2002). Further mission and instrumental details are described in Lin *et al*. (2002) and references therein.

### 2.1 Flare Selection

We restricted our analysis to M- and X-class flares – those most likely to produce super-hot plasma. To maximize the likelihood of observing the temperature peak, we required that the flare be well-observed, defined as uninterrupted coverage of the GOES SXR (1–8 Å) peak and the entire preceding 10-minute interval. We further required that the RHESSI HXR (25–50 keV) and SXR (6–12 keV) peaks be contained within this 10-minute interval and that they occur, in order, prior to the GOES SXR peak; the 10-minute length was chosen to include most flares while reducing extraneous processing.

To be able to compare all flares equally, we required that all time-series RHESSI spectra during the 10-minute analysis interval be acceptably fit by the spectral model described in §2.2. To ensure a reliable volume measurement, we required that selected flares could be successfully imaged using the methodology described in §2.3, at least around the time of the GOES SXR peak; two flares with clearly identifiable imaging artifacts were manually culled.

Given these criteria, 260 total flares – 234 M-class and 26 X-class – from 2002 to 2005 were appropriate for analysis; for this study, we selected a subset of 37 flares (25 M-class, 12 X-class), chosen in chronological order (Table 1). All selected M-class flares occurred in 2002, while the X-class flares, being less frequent, were chosen from 2002 to 2004 to ensure an adequate sample. The selected flares were distributed fairly randomly and uniformly in heliographic longitude (Figure 1, left).

### 2.2 Spectroscopy

Forward-modeling spectral analysis was performed using the *Object Spectral Executive* (OSPEX) package[1] in the *SolarSoft*[2]

[1] http://hesperia.gsfc.nasa.gov/rhessidatacenter/spectroscopy.html
[2] http://www.lmsal.com/solarsoft/

(SSW) IDL suite. Because of the extensive amount of data, the analysis was as largely automated as possible, though manually monitored at every step to ensure reliability.

For each selected flare, spectra were accumulated during the 10-minute observation period over the ~3–100 keV range with 1/3-keV energy binning (the instrument channel width) and 4-sec time binning (the spacecraft spin period); to maximize statistics, the spectra were averaged over all detectors except 2 and 7 (neither of which was usable for low-energy spectroscopy during this period). Successive time bins were summed into 20-second intervals, and the non-solar background was subtracted (see Caspi 2010 for details). Intervals spanning an attenuator-state transition (e.g., from thin-only to thick+thin) were ignored to prevent mixed-state observations, where the detector response is not well-defined.

At each 20-second interval, we used OSPEX to forward-fit a photon model (Figure 1, right) including a single isothermal continuum (using CHIANTI v5.2 [Landi *et al*. 2006] with coronal abundances), a non-thermal power-law continuum, and two Gaussian functions for the Fe and Fe–Ni lines (cf. Caspi & Lin 2010), convolved with the instrument response, to the observed spectrum above ~5.67 keV. The nominal calibration was used for the detector response (including pulse pileup), although observations in the thick+thin attenuator state were approximately corrected for a small but significant inaccuracy in the nominal thick attenuator response (see Caspi 2010) via the addition of a component to the photon model (Figure 1, right). The systematic uncertainty parameter in OSPEX was set to 2%, to account for uncharacterized discrepancies between the responses of the multiple detectors. For any interval, if the model fit failed to converge or if the best reduced $\chi^2$ exceeded 4.0, the flare was eliminated from consideration.

After achieving a reasonable fit at each interval, the interval with the largest fit isothermal continuum temperature was identified; focusing solely on this one interval per flare allows an equal comparison between flares, regardless of their duration or temporal variations. The maximum temperatures and cotemporal emission measures for all 37 flares are shown in Figure 2 and are listed in Table 1. For two of the 37 flares (2003 October 29 and 2003 November 03, both X-class events), the originally-identified maximum-temperature interval results were determined to be likely fitting artifacts as they resulted in unphysical temporal fluctuations of the temperature and emission measure, and exhibited clear systematics in the spectral fit residuals despite a reasonable $\chi^2$; for these two flares, these intervals were discarded and the maximum temperatures were identified from the remaining (acceptable) intervals.

### 2.3 Imaging

To estimate the volume occupied by the hot thermal plasma, for each flare, we created an image in the 6–15 keV range – which was, without exception, dominated by the thermal component – around the maximum-temperature time (40-sec duration), using subcollimators 3 through 9 (excluding 7) and the CLEAN image reconstruction algorithm with uniform weighting (Hurford *et al*. 2002). The thermal source volume was approximated by calculating the area $A$ within the 50% contour (see Figure 1, right, inset), corrected for broadening by the instrument point-spread function, and extrapolating to a volume by assuming spherical symmetry, $V = (4/3) \pi (A/\pi)^{3/2}$; this was done entirely automatically and, based on simulations, has an associated ~23% random uncertainty (see Caspi 2010 for full details), though with an unknown uncertainty related to projection of a 3D source onto a 2D image.

Assuming the volume to be roughly constant over the 10-minute observation interval, we combined the volume estimate $V$ with the time-series fit temperature $T$ and emission measure $Q$ to determine the thermal source density $n_e = \sqrt{Q/V}$, the total thermal energy (assuming $T_i = T_e$) $E_{th} = 3n_e V k_B T$, and the thermal energy density $E_{th}/V$, as well as propagated uncertainties, at both the time of maximum temperature and the time of maximum total energy. These measurements, for all 37 analyzed flares, are shown in Figures 3, 4, 5, and 6, and are listed in Table 1.

*2.4 Bias and Limitations*

Although our selection criteria may introduce some bias, the effects appear to be negligible. The requirement that a flare be successfully imaged with subcollimators 3–9 places an effective minimum threshold on source size of ~10″ FWHM, and requires that there be structure at this spatial scale even for larger sources – otherwise, the signal from the finer subcollimators will be noise-dominated, degrading the final image, particularly when using uniform weighting (as we do here). However, such noise artifacts were observed in only 2 out of 260 candidate flares, so this is not a significant effect. Our choice of only a 10-minute interval preceding the GOES SXR peak, which must also contain identifiable RHESSI HXR and SXR peaks in order, may exclude certain long-duration events that take a long time to cool from the initial HXR burst or ones that do not follow the standard flare heating/cooling model; however, none of the 260 flare candidates were excluded due to missing or "out of order" peaks, so these requirements appear justified.

We note that while the flares in our survey were generally fit well by a single isothermal component, the carefully-calibrated analysis of the 2002 July 23 X4.8 event (Caspi & Lin 2010) showed that, for that flare, two distinct isothermal components exist simultaneously throughout the flare. Multiple studies (e.g., Lin et al. 1981; Lin et al. 1985; Jakimiec et al. 1988; Longcope et al. 2010; Caspi & Lin 2010; Longcope & Guidoni 2011) suggest that double-isothermal distributions may be common in super-hot flares. Nevertheless, a spot-check comparison between the 2002 July 23 parameters derived here and by the more careful analysis of Caspi & Lin (2010) shows a ≲4% discrepancy in both the maximum temperature and cotemporal emission measure, suggesting that the omission of the lower-temperature component from the model fit does not significantly skew our results.

3. RESULTS AND INTERPRETATION

The primary results of our analysis are, in summary:
— The maximum RHESSI temperature is strongly correlated with GOES class, with a significantly steeper dependence than that of the GOES XRS-derived temperatures; and
— Super-hot ($T_e > 30$ MK) flares are strongly associated with higher thermal electron densities, energies, and energy densities compared to cooler ($T_e < 30$ MK) flares.

We present each of these results in more detail below, and discuss their implications for super-hot plasma in §4.

*3.1 Maximum Temperature*

Figure 2 (left) shows how the maximum RHESSI temperature $T_R$ varies with GOES class. Despite the large spread, there is a well-defined exponential relationship between the two quantities. The Pearson correlation coefficient $r$ between $T_R$ (in MK) and $\log_{10}$ of the GOES flux $F_G$ (in W m$^{-2}$) is ~0.88, with a fit relationship of $T_R \approx (14 \pm 1.2) \log_{10} F_G + (91 \pm 5.4)$.

The maximum GOES XRS isothermal temperatures $T_G$ were derived from the 3-second-cadence GOES SXR fluxes (pre-flare background subtracted) using the method of White et al. (2005), as implemented in the GOES workbench in SSW, and are also plotted, for comparison. (As with RHESSI, coronal abundances were assumed.) They, too, show a strong correlation, with $r \approx 0.84$, and an exponential fit relationship $T_G \approx (4.6 \pm 0.5) \log_{10} F_G + (38 \pm 2.2)$. This is within uncertainties of the relationship derived by Ryan et al. (2012) for over 52,000 GOES events and indicates that our 37 flares adequately sample the global population. We note that Ryan et al.'s slightly smaller slope of $3.9 \pm 0.5$ is heavily influenced by B- and C-class flares that dominate their sample population; restricting their population to only M- and X-class flares, as in this work, would further improve our agreement. (We also note that the GOES temperature maxima always occurred *after* the RHESSI maxima, which typically occurred at or just after the non-thermal HXR peak.)

The RHESSI temperatures are systematically higher than the GOES temperatures. However, a given flare is very likely *not* isothermal, but rather has a distribution of temperatures (e.g., McTiernan et al. 1999; Warren et al. 2013). RHESSI and GOES sample that temperature distribution differently – RHESSI is sensitive to temperatures above ~10 MK, with exponentially increasing sensitivity to hotter plasmas, while GOES is sensitive both to hot plasma (though less so than RHESSI) and to plasmas as cool as ~3–5 MK, which RHESSI cannot observe (cf. McTiernan 2009; Caspi & Lin 2010; Ryan et al. 2012). Given their different instrument responses, it is not surprising that RHESSI yields hotter emission-measure-weighted, isothermally approximated temperatures than does GOES, in general.

The relationship of the RHESSI temperatures to GOES class is ~3× steeper than that of the GOES temperatures, with >7σ confidence. Feldman et al. (1996), using BCS observations of S, Ca, and Fe excitation lines to determine plasma temperature for 868 flares of GOES class A2 to X2, found that $T_{BCS} \approx 5.4 \log_{10} F_G + 46$, only somewhat steeper than our GOES relationship $T_G$. Because they chose the temperature from the GOES SXR peak time, whence the temperature is ~10% cooler than the actual temperature peak (Ryan et al. 2012), their true correlation could be somewhat steeper by this same factor, though still far shallower than our RHESSI-observed behavior $T_R$. However, due to instrument saturation during intense flares, their correlation may be questionable for flares above GOES class M2. Battaglia et al. (2005) and Hannah et al. (2008), using RHESSI data and analysis techniques similar to ours, also found shallower correlations, with Battaglia et al. (2005) reporting $T_R \approx 3.0 \log_{10} F_G + 35$ (albeit with near 90% uncertainty in the fit parameters due to the large scatter in their data) for 85 flares from GOES class B1 to M6; Hannah et al. (2008) did not quantify their correlation of over 25,000 microflares up to GOES class of ~C3, but indicated a slope equal to or shallower than that of Battaglia et al. (2005).

Over the M- and X-class range studied here, the ratio $T_R/T_G$ increases from ~1.3 (at M1) to ~2.1 (at X10), yielding a linearly approximated relationship of $T_R \approx 3.2 T_G - 28.6$ over this range ($15.5 \lesssim T_G \lesssim 24.5$). For less intense flares, much shallower correlations have been reported: Battaglia et al. (2005) found, for GOES classes B1–M6, $T_R \approx 1.12 T_G + 3.12$ with a range of $4 \lesssim T_G \lesssim 20$; Hannah et al. (2008) did not give a specific correlation but their analysis of GOES class <C3 implies a slope significantly smaller than unity ($T_R \approx 0.56 T_G + 6.2$, based on an eyeball fit to their Figure 14 [top]) with the same range of $T_G$. Our data, with increasing $T_R/T_G$ versus GOES class, indicates that hotter temperatures increase preferentially, compared to cooler ones, with increasing flare intensity, while Hannah *et al.* (2008) indicates the opposite behavior for less intense flares.

Figure 2 (right) shows the relationship between $T_R$ and its corresponding (cotemporal) emission measure. If our high temperatures were an artifact of the fitting process, as is sometimes observed, we would expect the temperature to be anti-correlated with emission measure; this is not the case, indicating that our temperature variations are likely real, barring other sources of systematic error. While there is no significant correlation between the temperature and emission measure ($r \approx 0.30$ in log–linear space), there is nevertheless an apparent threshold association, with 13 of 14 super-hot flares having an emission measure exceeding $\sim 4.0 \times 10^{47}$ cm$^{-3}$, and 10 of 14 exceeding $\sim 1.3 \times 10^{48}$ cm$^{-3}$; in contrast, the cooler flares span the entire range of emission measures. Although these results are not directly comparable with those of Feldman *et al.* (1996), as their emission measures were derived at the GOES peak rather than at the temperature maximum, our observations of high emission measures for super-hot flares nevertheless suggest that Feldman *et al.*'s conclusions – that such flares must scale intrinsically and cannot be the sum of many small, unresolved events – are also applicable here.

### 3.2 Volume and Density

Figure 3 shows how the derived thermal source volume and density vary with maximum temperature. The thermal source volume shows no correlation ($r \approx 0.10$ log–linear) and is evenly distributed; this suggests that there is no preferred macroscopic ("global") physical size for super-hot versus cooler flares. While density also shows no specific correlation ($r \approx 0.15$ log–linear), it is not evenly distributed – as with the emission measure, the density shows a strong threshold association, with 12 of 14 super-hot flares exceeding $\sim 3.2 \times 10^{10}$ cm$^{-3}$, while the cooler flare densities span the entire range. (The outliers are primarily associated with questionably-large volume measurements, where the images show a complex morphology and multiple sources, thus invalidating the "isothermal single source" assumption for the density calculation.) This suggests a potential minimum density threshold for the formation of super-hot plasma; such a threshold appears necessary, but not sufficient, as cooler flares can also exhibit high densities.

We note that these densities are actually stringent lower limits, as we assumed a volume filling factor of unity; if the RHESSI images do not fully resolve any existing fine structure, the true filling factor $f$ may be smaller, and since $n_e \propto f^{-1/2}$, the density would be correspondingly larger. However, physical arguments provide a lower bound for $f$ – for super-hot flares, the densities are already high assuming $f \approx 1$; as $f$ decreases, $n_e$ quickly approaches chromospheric values. For physically plausible values of $n_e \lesssim 10^{12}$ cm$^{-3}$, $f$ must be no smaller than $\sim 0.01$.

### 3.3 Energy

Figure 4 shows the thermal energy at the time of, and versus, the maximum RHESSI temperature. The total thermal energy (Figure 4, left), assuming $T_i = T_e$, shows a moderate correlation ($r \approx 0.53$ log–linear), but a strong threshold association, with 13 of 14 super-hot flares exceeding $\sim 2.4 \times 10^{29}$ erg in the thermal plasma at the time of the temperature maximum, while cooler flares vary across a wide range. The thermal energy *density* (Figure 4, right) shows a weaker correlation ($r \approx 0.45$) but a similarly strong association, with 13 of 14 super-hot flares exceeding $\sim 450$ erg cm$^{-3}$. As with the electron number densities, the measured energy densities are strict lower limits due to the assumed unity filling factor.

The minimum threshold associations are even more strongly observed for the maximum thermal energy, which occurs later in the flare than the maximum temperature; Figure 5 shows the maximum energy and associated energy density versus GOES class, while Figure 6 shows these values versus the maximum RHESSI temperature achieved (earlier) during the flare. A strong correlation ($r \approx 0.83$ in log–log space) is observed between the maximum thermal energy $E_{\max}$ and the GOES flux $F_G$, with a fit relationship of $\log_{10} E_{\max} \approx (0.64 \pm 0.074) \log_{10} F_G + (32 \pm 0.33)$. When compared against maximum temperature, a striking association is observed – *none* of the non-super-hot ($T < 30$ MK) flares have a maximum thermal energy beyond $\sim 9 \times 10^{29}$ erg, while 9 of 14 super-hot flares exceed this value. The thermal energy *density* also shows both a strong correlation ($r \approx 0.72$ log–log for GOES class, and $r \approx 0.70$ log–linear for temperature) and a strong threshold association, with 11 of 12 X-class flares, and 12 of 14 super-hot flares, exceeding $\sim 1300$ erg cm$^{-3}$, while weaker/cooler flares vary widely with no apparent threshold. (As above, the low-lying super-hot outliers are primarily associated with potential "multiple source" images.)

The association of super-hot flares with high total thermal energies, both early in the flare, at the time of the maximum temperature (Figure 4), and at the subsequent energy maximum (Figures 5 and 6), suggests that super-hot flares require a greater overall energy input compared to cooler flares. This is particularly evident when considering that the energy measurements presented here are *instantaneous*, and do not account for radiative or conductive losses. At temperatures $\gtrsim 20$ MK, radiation is dominated by continuum emission (free–free and free–bound), and the radiative loss function increases monotonically with both density and temperature; in addition to the obvious fact that brighter (i.e., more intense) flares, by definition, radiate more energy, the higher temperatures and (on average) densities of super-hot flares imply that they have still larger overall radiative losses. Thus, the higher instantaneous thermal energies require an even-greater input of energy to the thermal plasma in super-hot flares, compared to cooler flares.

Thermal energy density is equivalent to plasma kinetic pressure, and the high values observed for super-hot flares, both at the time of the maximum temperature (Figure 4) and at the later energy maximum (Figures 5 and 6), become even more intriguing when comparing them to the magnetic pressure. If we consider the plasma $\beta$ for an isothermal source contained by magnetic fields, then we require that $\beta$ never exceed 1 – if it did, the plasma kinetic pressure would dominate the magnetic pressure and could push the fields apart, allowing the plasma to expand and cool adiabatically; if $\beta < 1$, the field pressure dominates and it can keep the plasma confined, preventing it from cooling by expansion.[3] Thus, *the measured thermal energy density corresponds to the minimum field strength required to contain the plasma* – this is represented by the horizontal dotted lines in Figures 4, 5, and 6 (right). Thirteen of 14 super-hot flares require a *coronal* field strength exceeding $\sim 100$ G at the time of the temperature maximum, or $\sim 160$ G at the energy maximum. This suggests that a minimum threshold exists for the magnetic field strength, below which a super-hot plasma cannot form; as with the number density, this threshold condition appears necessary, but not sufficient, as strong fields may exist in cooler flares, as well.

---

[3] Although this pressure-balance argument assumes a static scenario, while our measurements are snapshots of a dynamic process, adiabatic expansion of the plasma would occur at roughly the sound speed, $c_s \equiv \sqrt{\gamma k_B T / \bar{\mu} m_p}$, where the adiabatic index $\gamma = 5/3$, $m_p$ is the proton mass, and the average molecular mass $\bar{\mu} \approx 0.6$ in the solar corona; for a plasma of average radius $r \approx 5 \times 10^3$ km and average temperature $T \approx 30$ MK, as we have in our sample, the characteristic expansion time $\tau_a \equiv r/c_s \approx 6$ s, far shorter than our spectral analysis time of 20 s. Additionally, an expanding plasma would exhibit a decrease in density after the temperature maximum, opposite to what we observe. For both reasons, the quasi-static approximation appears valid here.

We note that since any inferred $B \propto f^{-1/4} \beta^{-1/2}$, the inferred $B$ values are a strict lower limit, as both $\beta$ and the volume filling factor $f$ must be unity or smaller. However, in the region of the super-hot plasma – at the top of coronal loops – $B$ may not physically exceed the field strength in the chromosphere or photosphere, which is typically no more than ~1000–3000 G, i.e., no more than ~10× our *minimum* inferred coronal field, even for the most intense active regions (e.g., White *et al.* 1991). Such physical upper bounds on $B$ place strict lower limits on $\beta$ in the super-hot region, requiring that $1 \lesssim f^{-1/4} \beta^{-1/2} \lesssim 10$ and hence $\beta \gtrsim$ ~0.01. Indeed, radio observations (e.g., Asai *et al.* 2006, for the 2002 July 23 X4.8 flare) suggest looptop field strengths of only a few hundred gauss, consistent with our inferred field values, suggesting (with §3.2) that both $\beta$ and $f$ are not much less than unity. We note that Krucker *et al.* (2010), using radio observations of an M2 event to measure a field strength of ~50 G in a flaring region high in the corona, also reported $\beta \approx 1$ for their event, although their observations were of non-thermal electrons.

## 4. DISCUSSION

Our survey has revealed an intriguing correlation between GOES class and the RHESSI-observed maximum temperature $T_R$ that differs significantly from a similar correlation with GOES XRS-derived temperatures $T_G$. A strong correlation was also observed between GOES class and the maximum RHESSI-measured thermal energy. What implications do these results have for the origins of super-hot plasma?

The ~10–20 MK flare plasma that dominates the GOES response is widely considered to result from evaporation of chromospheric material heated by the collisional energy losses from accelerated, non-thermal electrons impacting the ambient medium (e.g., Fletcher *et al.* 2011; Holman *et al.* 2011). Previous studies have shown that GOES class has a positive correlation with the HXR instantaneous flux (Battaglia *et al.* 2005) and flare-integrated fluence (Veronig *et al.* 2002), which are diagnostics of the non-thermal electron population. Although the HXR flux/fluence is not a direct measure of the *energy* contained in non-thermal electrons – knowledge of the spectral index and low-energy "cutoff" would also be required – the evaporative origins of the GOES plasma and correlation with HXR flux strongly suggest that GOES class could be a reasonable proxy for the energy deposited into the chromosphere by non-thermal electrons (viz. the Neupert effect – Neupert 1968; Dennis & Zarro 1993), and therefore, indirectly (though perhaps more loosely), for the energy released in the acceleration region via magnetic reconnection.

Assuming GOES class is such a proxy, then, if the super-hot plasma that dominates the RHESSI response also resulted from chromospheric evaporation, one would expect that the relationship between maximum temperature and GOES flux (or, by proxy, energy deposition) would behave approximately the same, at least in terms of slope, for RHESSI-observed temperatures as for GOES-observed ones; the significant discrepancy between our RHESSI and GOES correlations thus suggests that super-hot plasmas are formed via a different physical process, one which depends more sensitively on the amount of energy released during the flare. Indeed, no numerical simulations of chromospheric evaporation from collisional energy losses (e.g., Fisher *et al.* 1985; Allred *et al.* 2005) have yet been able to reproduce super-hot temperatures using physically realistic inputs, thus supporting a different physical origin. The relative timing of the temperature maxima – near the HXR peak for RHESSI, and later for GOES – is also consistent with an *in situ* mechanism for super-hot plasma, which can occur nearly simultaneously with energy release, and a transport mechanism for cooler plasma, which is somewhat delayed.

Caspi & Lin (2010) proposed that, during the pre-impulsive phase of an X-class flare, the super-hot plasma is formed via compression (and subsequent thermalization) of reconnection-outflow material by the magnetic fields of the reconnecting flux ropes as they relax into more potential configurations; the results here suggest that this proposed mechanism is likely applicable even during the impulsive phase. The gas dynamic shock formation mechanism proposed by Longcope & Guidoni (2011), which also yields super-hot temperatures and density enhancements, would also be consistent, as might be other mechanisms as yet unexplored. Importantly, if one extends the RHESSI- and GOES-derived fit temperature functions ($T_R$ and $T_G$, respectively) to lower GOES fluxes ($F_G$), they cross at GOES class of ~C4, suggesting that, whatever the *in situ* heating process might be, it is not limited to super-hot flares but is present even in less intense flares that do not achieve super-hot temperatures. The power-law correlation between maximum thermal energy and GOES class hints at an intimate connection between the *in situ*-heated plasma and the energy released via reconnection (by proxy with GOES class).

Our results further show that super-hot flares are strongly associated with high densities, compared to cooler flares. This suggests a potential minimum density threshold for the formation of super-hot plasma. This would be consistent with the super-hot formation mechanisms of Caspi & Lin (2010) and Longcope & Guidoni (2011), wherein higher densities would allow the reconnection outflow to thermalize more quickly *at the looptop*, while lower densities would yield longer thermalization times, allowing some or all of the outflow material to remain non-thermal and escape to the footpoints, where it would contribute to the formation of cooler plasma via chromospheric heating and evaporation (see also Sakao *et al.* 1998). However, other explanations for a high density are also possible.

The question remains, then: are super-hot flares merely hotter versions of cooler flares, or a separate class of event? The RHESSI temperature correlation extends below super-hot temperatures, down to flares as weak as C9 in our observations (and potentially C4, per above), and thus the *in situ* coronal heating likely exists even for flares where the directly-heated plasma does not reach super-hot temperatures. Indeed, this could explain the large scatter we observe in the emission measure, density, energy, and energy density for cooler, non-super-hot flares where the directly-heated and chromospherically-evaporated plasma temperatures become similar – in such flares, the RHESSI measurements are influenced by *both* plasmas, whereas in super-hot flares, they are dominated by the (much hotter) *in situ*-heated plasma alone. On the other hand, if the two spatially-distinct plasmas influence RHESSI equally, one might expect that the measured volumes would be, on average, larger for cooler flares than for super-hot, or at least that the images would show multiple sources more often for the cooler flares, but we observe neither – the volumes appear dominated by single sources and the average volume is identical for super-hot and cooler flares. Although not trivial, this could potentially be addressed via imaging spectroscopy for these cooler flares. Nevertheless, it is still unclear whether super-hot flares truly are a separate class of event.

It does appear likely, however, that flares where *in situ* heating actually occurs may be distinct. Compared to the GOES temperatures, we derive a steeper correlation of RHESSI temperatures with GOES class for flares down to ∼C4, while Battaglia *et al.* (2005) and Hannah *et al.* (2008) observe a shallower correlation for flares below ∼C3 down to sub-A class. The evolution of the temperature distribution thus seems to change abruptly around the C4 level. Again assuming GOES class as a proxy for energy release, this could suggest a potential energy threshold, below which *in situ* coronal heating is impeded or nonexistent, but above which it proceeds efficiently, scaling with the amount of energy released. It would be instructive to examine how the peak RHESSI temperature, density, etc. correlate with non-thermal *electron* flux/fluence (versus the HXR *photon* flux/fluence in previous studies), or more precisely, with the non-thermal *energy* flux/fluence, and to compare these measurements with hydrodynamic modeling of the atmospheric response (e.g. Allred *et al.* 2005) to the observed non-thermal parameters; this is the subject of a future study.

## 5. SUMMARY AND CONCLUSIONS

We have used RHESSI to accurately determine the peak electron temperature and associated emission measure, thermal electron density, energy, and energy density in 37 M- and X-class flares. We have shown that the maximum achieved RHESSI temperature is strongly correlated with GOES class, and far more steeply than is the GOES XRS-derived temperature. We also determined that super-hot flares are strongly associated with high densities, thermal energies, and energy densities. All of these results support the concept of the RHESSI-observed super-hot plasma being heated directly in the corona, a physically distinct process from the chromospheric evaporation that creates the GOES-temperature plasma.

Although the 30 MK threshold for the "super-hot" moniker appears arbitrary, it has a physical significance: the chromospherically-evaporated plasma does not seem to breach this temperature, while the directly-heated plasma temperature rises steeply, possibly limited only by the energy content of the flare. Super-hot plasmas are thus a direct probe of the *in situ* heating process, and further observations could help determine what specific physical mechanism dominates the heating.

Our measurements of the thermal energy density during both the peak-temperature interval and the peak-energy interval reveal that super-hot flares are ubiquitously associated both with high total energies and with strong ($\gtrsim$100 G) inferred magnetic fields in the corona; this suggests that a minimum field strength is a necessary, though perhaps not sufficient, condition for the formation of super-hot plasma, and that super-hot flares reflect not only higher temperatures, but higher actual energy input into the thermal plasma. If the formation of super-hot plasma is tied to the reconnection process, as suggested by Caspi & Lin (2010) and Longcope & Guidoni (2011), this may help to distinguish between various reconnection models for flares that achieve super-hot temperatures and/or exhibit *in situ* heating in general, versus those that do not. In concert with radio observations, the inferred field strengths constrain the plasma $\beta$ and filling factor to near unity within the super-hot flaring region, which suggests that the plasma is being efficiently heated to its physical maximum and is an important consideration when considering dynamical effects (e.g., wave propagation) in the flaring loop.

This manuscript is dedicated to the memory of R. P. Lin, whose vision, inspiration, and enthusiasm will be sorely missed. This work was supported in part by NASA contract NAS5-98033. A. Caspi was also supported in part by NASA contract NAS5-02140 and NASA grants NNX08AJ18G and NNX12AH48G, and R. P. Lin by the WCU grant (No. R31-10016) funded by the Korean Ministry of Education, Science and Technology. The authors thank A. G. Emslie, H. S. Hudson, R. A. Schwartz, and especially A. Y. Shih for helpful discussions.

**Table 1**
Measured and Derived Plasma Parameters for the Surveyed Flares

| Date | GOES Class | $T_G$ (MK) | Time[a] (UT) | $T_R$ (MK) | EM[b] ($10^{47}$ cm$^{-3}$) | $V$[b] ($10^{26}$ cm$^3$) | $n_e$[b] ($10^{10}$ cm$^{-3}$) | $E_{th}$[b] ($10^{28}$ erg) | $E_{th}/V$[b] (erg cm$^{-3}$) | $B_{\beta=1}$[b] (G) | $E_{th}$ (max.)[c] ($10^{28}$ erg) | $E_{th}/V$ (max.)[c] (erg cm$^{-3}$) | $B_{\beta=1}$ (max.)[c] (G) |
|---|---|---|---|---|---|---|---|---|---|---|---|---|---|
| 2002 Feb 20 | M4.3 | 17 | 09:57:50 | 25 | 55 | 1.7 | 18 | 32 | 1900 | 220 | 35 | 2100 | 230 |
| 2002 Feb 20 | M2.4 | 15 | 21:06:10 | 21 | 6.4 | 0.84 | 8.7 | 6.4 | 770 | 140 | 13 | 1500 | 200 |
| 2002 Feb 22 | M4.4 | 16 | 00:00:30 | 20 | 56 | 13 | 6.6 | 71 | 560 | 120 | 85 | 670 | 130 |
| 2002 Mar 17 | M1.3 | 16 | 10:15:30 | 23 | 4.2 | 7.2‡ | 2.4‡ | 16‡ | 230‡ | 76‡ | 33‡ | 470‡ | 110‡ |
| 2002 Apr 04 | M1.4 | 21 | 10:44:50 | 24 | 11 | 0.54 | 14 | 7.8 | 1400 | 190 | 11 | 2100 | 230 |
| 2002 Apr 07 | C9.6 | 13 | 02:26:50 | 20 | 5.0 | 14 | 1.9 | 21 | 160 | 63 | 28 | 200 | 72 |
| 2002 Apr 10 | M1.6 | 14 | 19:02:50 | 26† | 1.4† | 9.4‡ | 1.2‡ | 12‡ | 130‡ | 57‡ | 45‡ | 480‡ | 110‡ |
| 2002 Apr 11 | C9.2 | 13 | 16:19:50 | 19 | 1.2 | 4.4 | 1.7 | 5.8 | 130 | 58 | 19 | 440 | 110 |
| 2002 Apr 12 | M4.0 | 16 | 17:57:10 | 23 | 61 | 6.0 | 10 | 57 | 950 | 160 | 61 | 1000 | 160 |
| 2002 Apr 15 | M3.7 | 16 | 00:11:30 | 24† | 46† | 11.5 | 6.3 | 71 | 620 | 120 | 78 | 680 | 130 |
| 2002 Apr 16 | M2.5 | 16 | 13:10:50 | 22 | 21 | 6.3 | 5.8 | 34 | 540 | 120 | 50 | 800 | 140 |
| 2002 Apr 17 | C9.8 | 14 | 16:54:30 | 19 | 2.4 | 0.42 | 7.5 | 2.5 | 580 | 120 | 6.3 | 1500 | 190 |
| 2002 Apr 24 | M1.7 | 17 | 21:54:50 | 30 | 7.0 | 1.9 | 6.1 | 14 | 750 | 140 | 23 | 1200 | 180 |
| 2002 May 04 | C9.3 | 14 | 13:19:30 | 22† | 3.4† | 5.7‡ | 2.4‡ | 12‡ | 220‡ | 74‡ | 22‡ | 380‡ | 100‡ |
| 2002 May 20 | M5.0 | 19 | 10:52:30 | 29 | 37 | 0.81 | 21 | 21 | 2600 | 260 | 21 | 2600 | 260 |
| 2002 May 27 | M2.0 | 15 | 18:04:50 | 34 | 1.6 | 1.5 | 3.2 | 7.0 | 460 | 110 | 21 | 1400 | 190 |
| 2002 Jul 03 | X1.5 | 20 | 02:10:30 | 46 | 15 | 1.0 | 12 | 24 | 2300 | 240 | 48 | 4600 | 340 |
| 2002 Jul 04 | M1.1 | 16 | 07:31:50 | 21 | 1.8 | 10 | 1.3 | 12 | 120 | 55 | 20 | 190 | 69 |
| 2002 Jul 08 | M1.6 | 17 | 09:17:50 | 24 | 21 | 4.7 | 6.6 | 31 | 650 | 130 | 35 | 750 | 140 |
| 2002 Jul 18 | M2.2 | 20 | 03:34:10 | 32 | 16 | 3.4 | 6.9 | 30 | 910 | 150 | 33 | 970 | 160 |
| 2002 Jul 23 | X4.8 | 24 | 00:28:10 | 42 | 84 | 3.5 | 15 | 95 | 2700 | 260 | 160 | 4400 | 330 |
| 2002 Jul 26 | M1.0 | 14 | 18:59:50 | 21 | 3.1 | 1.6 | 4.3 | 6.3 | 380 | 98 | 13 | 770 | 140 |
| 2002 Jul 31 | C9.6 | 16 | 09:54:50 | 24 | 3.9 | 4.5 | 2.9 | 13 | 290 | 86 | 25 | 550 | 120 |
| 2002 Aug 03 | X1.0 | 19 | 19:06:10 | 34 | 36 | 7.9 | 6.8 | 75 | 950 | 160 | 100 | 1300 | 180 |
| 2002 Aug 11 | C9.5 | 17 | 11:40:50 | 26 | 2.3 | 44‡ | 0.73‡ | 34‡ | 79‡ | 45‡ | 78‡ | 180‡ | 67‡ |
| 2002 Sep 06 | C9.2 | 17 | 16:27:10 | 23 | 12 | 3.1 | 6.2 | 18 | 580 | 120 | 20 | 660 | 130 |
| 2002 Sep 09 | M2.1 | 15 | 17:45:30 | 26 | 5.7 | 3.4 | 4.1 | 15 | 430 | 100 | 30 | 860 | 150 |
| 2002 Sep 27 | C9.9 | 13 | 03:35:50 | 19 | 11 | 4.3 | 5.1 | 17 | 390 | 99 | 21 | 490 | 110 |
| 2002 Oct 31 | X1.2 | 22 | 16:51:30 | 39 | 32 | 0.84 | 20 | 26 | 3200 | 280 | 26 | 3200 | 280 |
| 2003 Jun 10 | X1.3 | 20 | 23:58:50 | 37 | 20 | 2.7 | 8.6 | 36 | 1300 | 180 | 67 | 2500 | 250 |
| 2003 Oct 19 | X1.1 | 21 | 16:40:30 | 32 | 55 | 8.0‡ | 8.3‡ | 89‡ | 1100‡ | 170‡ | 110‡ | 1300‡ | 180‡ |
| 2003 Oct 29 | X10. | 23 | 20:41:10 | 53† | 13† | 23‡ | 2.4‡ | 120‡ | 520‡ | 120‡ | 570‡ | 2500‡ | 250‡ |
| 2003 Nov 02 | X8.3 | 23 | 17:16:50 | 44 | 88 | 76‡ | 3.4‡ | 470‡ | 620‡ | 120‡ | 1100‡ | 1400‡ | 190‡ |
| 2003 Nov 03 | X2.7 | 22 | 01:20:30 | 37 | 96 | 4.3 | 15 | 98 | 2300 | 240 | 150 | 3400 | 290 |
| 2003 Nov 03 | X3.9 | 22 | 09:48:10 | 53 | 5.8 | 2.1 | 5.3 | 24 | 1200 | 170 | 110 | 5400 | 370 |
| 2004 Feb 26 | X1.1 | 19 | 01:55:10 | 38† | 4.0† | 45‡ | 0.94‡ | 67‡ | 150‡ | 61‡ | 270‡ | 590‡ | 120‡ |
| 2004 Jul 15 | X1.8 | 24 | 01:37:50 | 38 | 7.9 | 3.6 | 4.7 | 26 | 730 | 140 | 130 | 3600 | 300 |

**Notes.**
[a] Quoted times represent the center of the maximum-RHESSI-temperature interval; the maximum GOES temperature $T_G$ occurs later in the flare.
[b] Measured/derived quantities are quoted for the time of the maximum RHESSI temperature, $T_R$.
[c] Quantities derived using volume at maximum RHESSI temperature (but spectral fits at maximum RHESSI energy)
[†] Spectral model fit yielded reduced $\chi^2 > 2$.
[‡] Images show complex morphology with likely multiple sources and significant energy dependence, invalidating the "isothermal single source" assumption. The volume is therefore most likely an overestimate (by an unknown factor), as is the corresponding total energy, while the corresponding number and energy densities are likely underestimated.

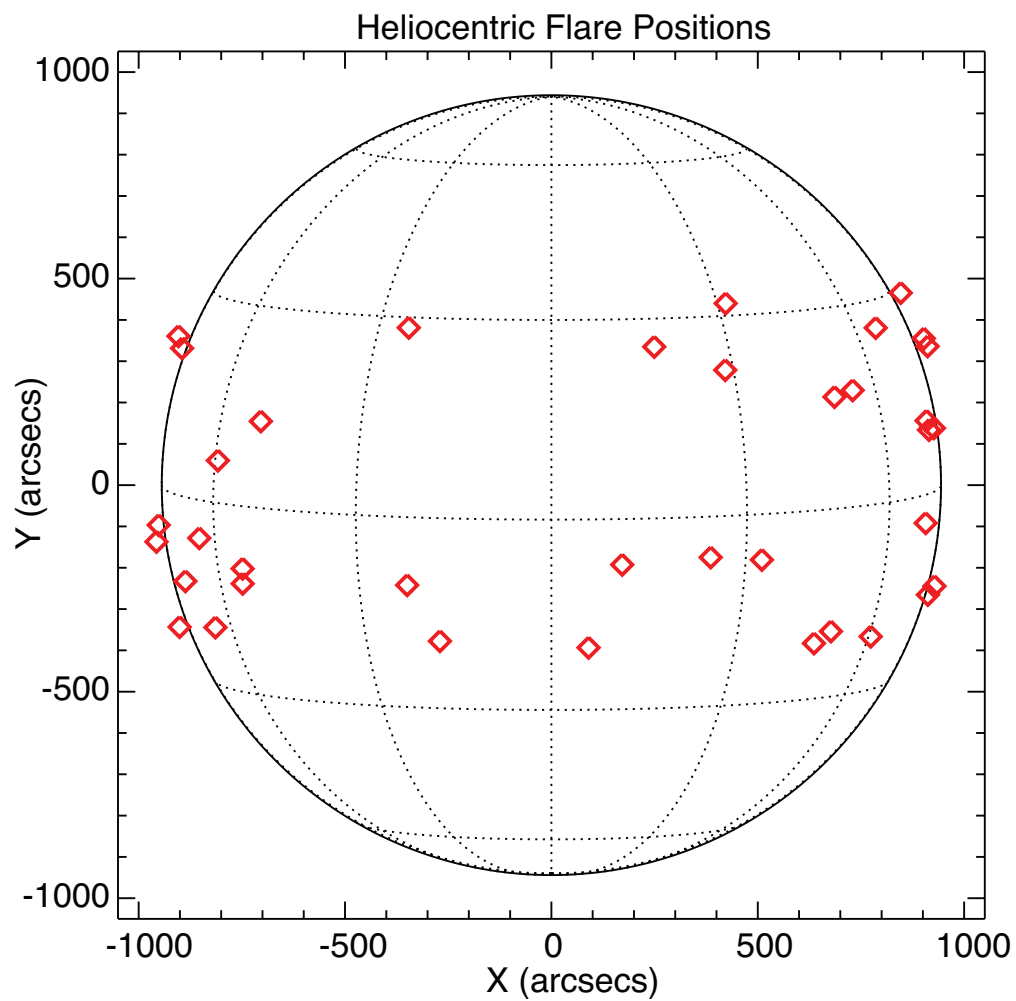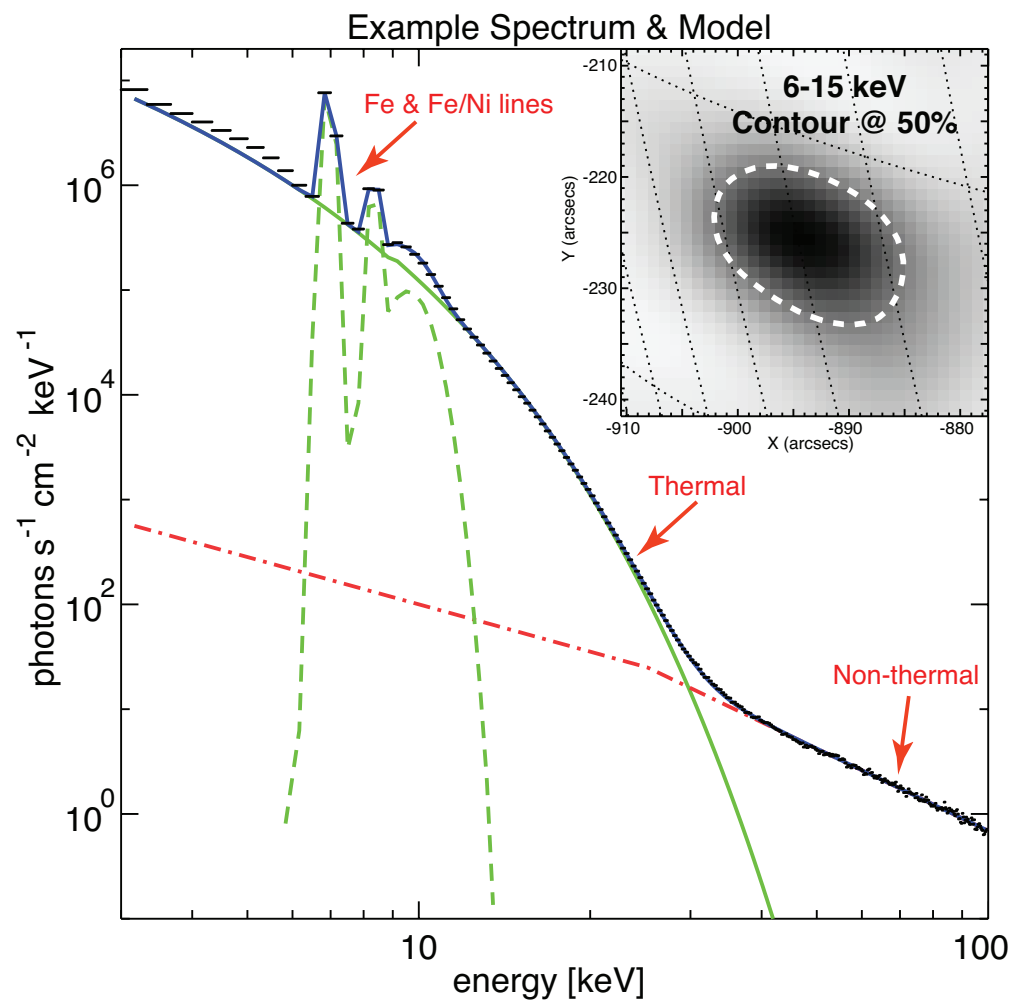

**Figure 1.** [*left*] Synoptic map of heliocentric positions for the 37 selected flares. [*right*] Example photon model used for spectral forward-fitting, including an isothermal continuum (*solid*), a non-thermal power-law continuum (*dot-dashed*), and two Gaussians representing the Fe and Fe–Ni unresolved line complexes (*dashed*). For observations in the thick+thin attenuator state, a third, wide Gaussian was added to account for an inaccuracy in the calibrated response of the thick attenuator. The 6–15 keV image (*inset; reverse color*) was used to estimate the thermal source volume from the area enclosed by the 50% brightness contour, corrected for broadening from the instrument point-spread function.

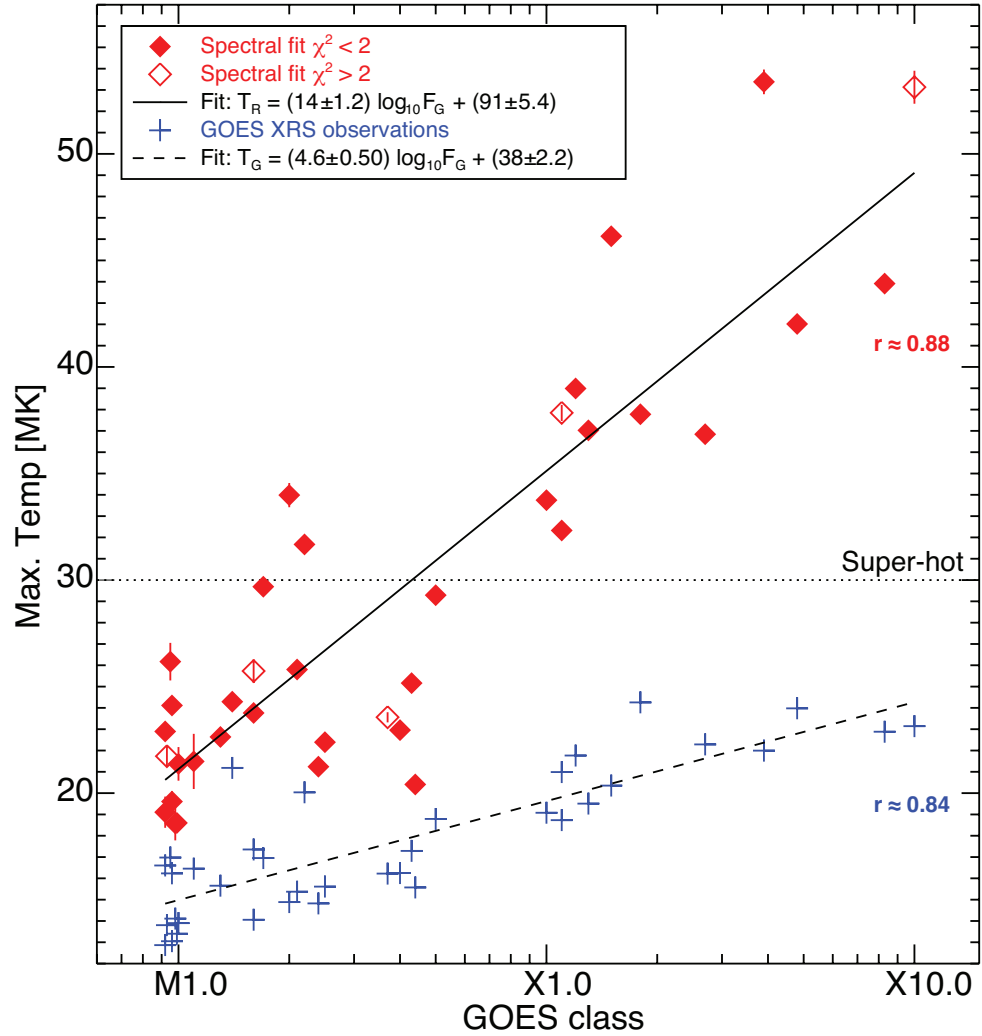
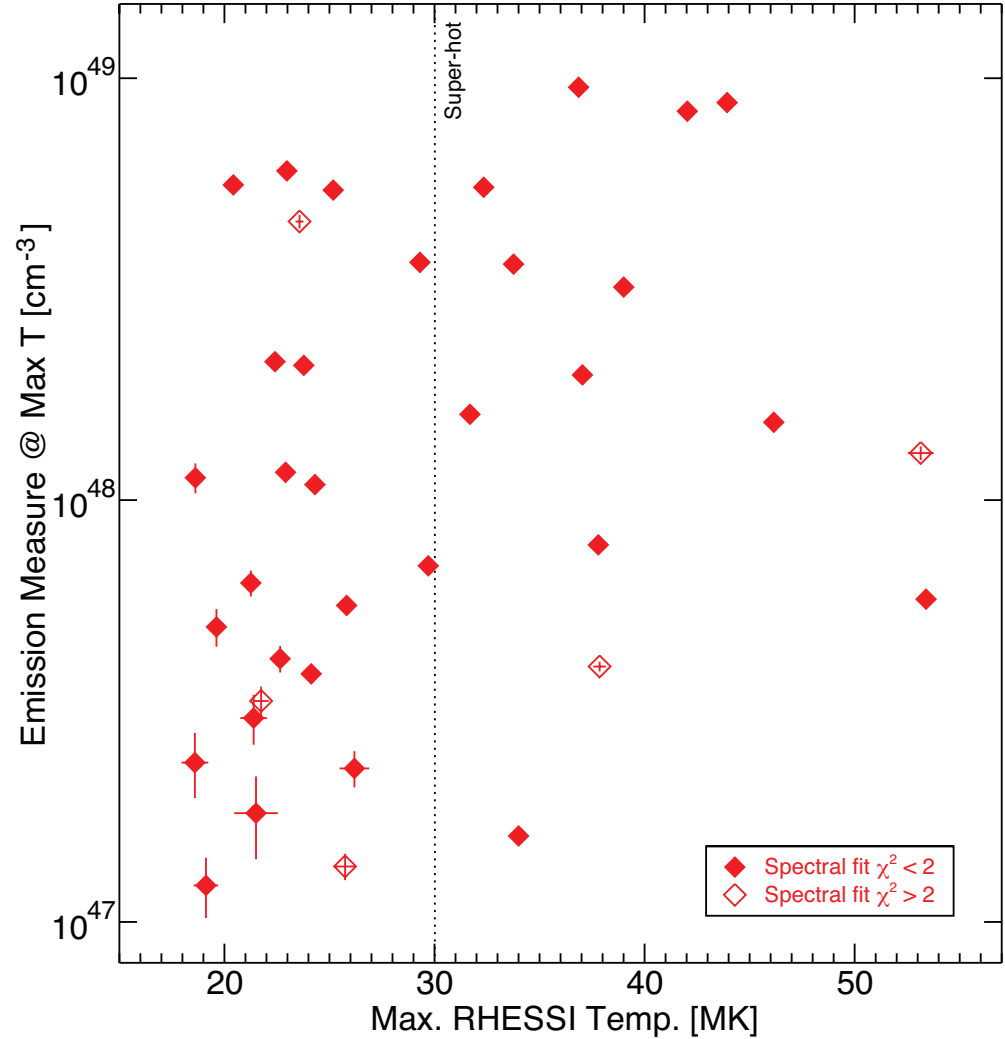

**Figure 2.** [*left*] Maximum RHESSI-measured isothermal continuum temperature (*diamonds*) vs. GOES class for the 37 analyzed flares, with fit correlation. Spectral fits with a reduced $\chi^2 > 2$ are denoted by *open* diamonds; they are distributed evenly in GOES class and thus do not significantly skew the observed correlation. All 12 X-class flares, but only 2 of 25 M-class flares, achieve super-hot (>30 MK) temperatures. *Plusses* denote the peak GOES XRS-derived isothermal temperatures for the same flares. [*right*] Emission measure corresponding to, and vs., the maximum measured continuum temperature. Thirteen of 14 super-hot flares have an emission measure exceeding $\sim 4 \times 10^{47}$ cm$^{-3}$. (The outlier is a limb flare on 2002 May 27 and was a "failed eruption" event, cf. Ji et al. 2003.)

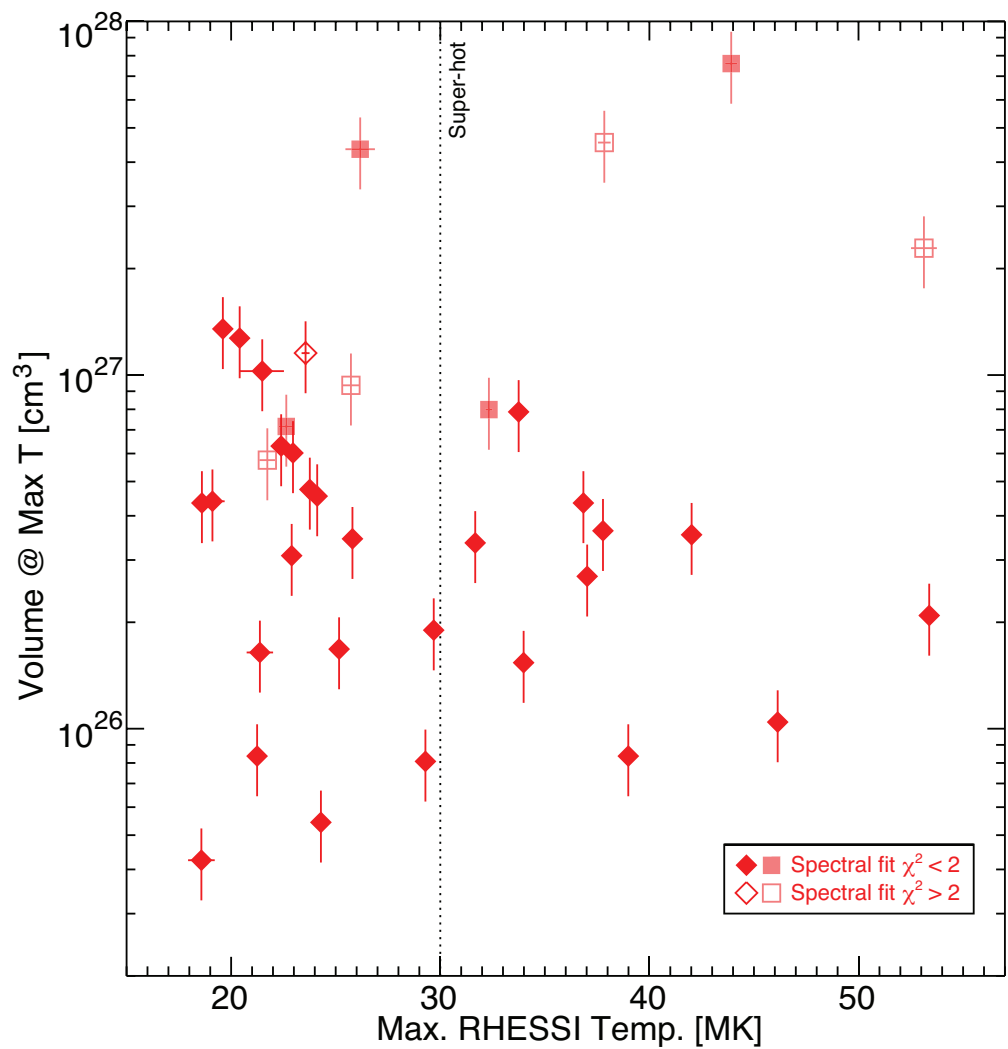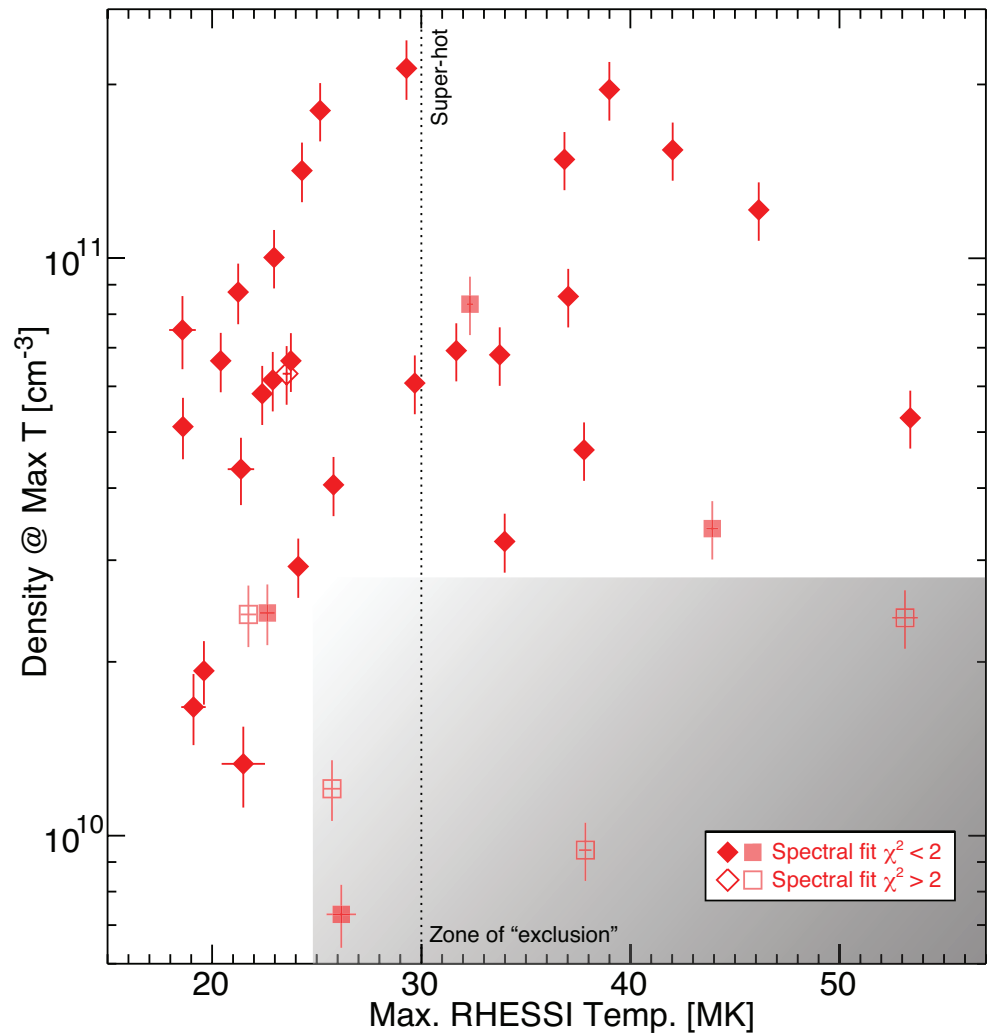

**Figure 3.** [*left*] Estimated volume derived from the 6–15 keV images cotemporal with, and vs., the maximum measured continuum temperature. The distribution is roughly uniform. In a few cases (*square* symbols), the images show a complex morphology and suggest that multiple sources may be present, skewing the volume measurement, which assumes only a single source; note that the four largest volumes all suffer from this issue, and two of those also exhibit poor $\chi^2$ values (*open* symbols) for the spectral fit. [*right*] Electron density cotemporal with, and vs., the maximum measured continuum temperature. Twelve of 14 super-hot flares have a density exceeding $\sim 3.2 \times 10^{10}$ cm$^{-3}$; lower densities appear to be "excluded" for super-hot flares, illustrated qualitatively by the "zone of exclusion" (*shaded*). The outliers are associated with the uncertain "multiple source" volume measurements (*squares*).

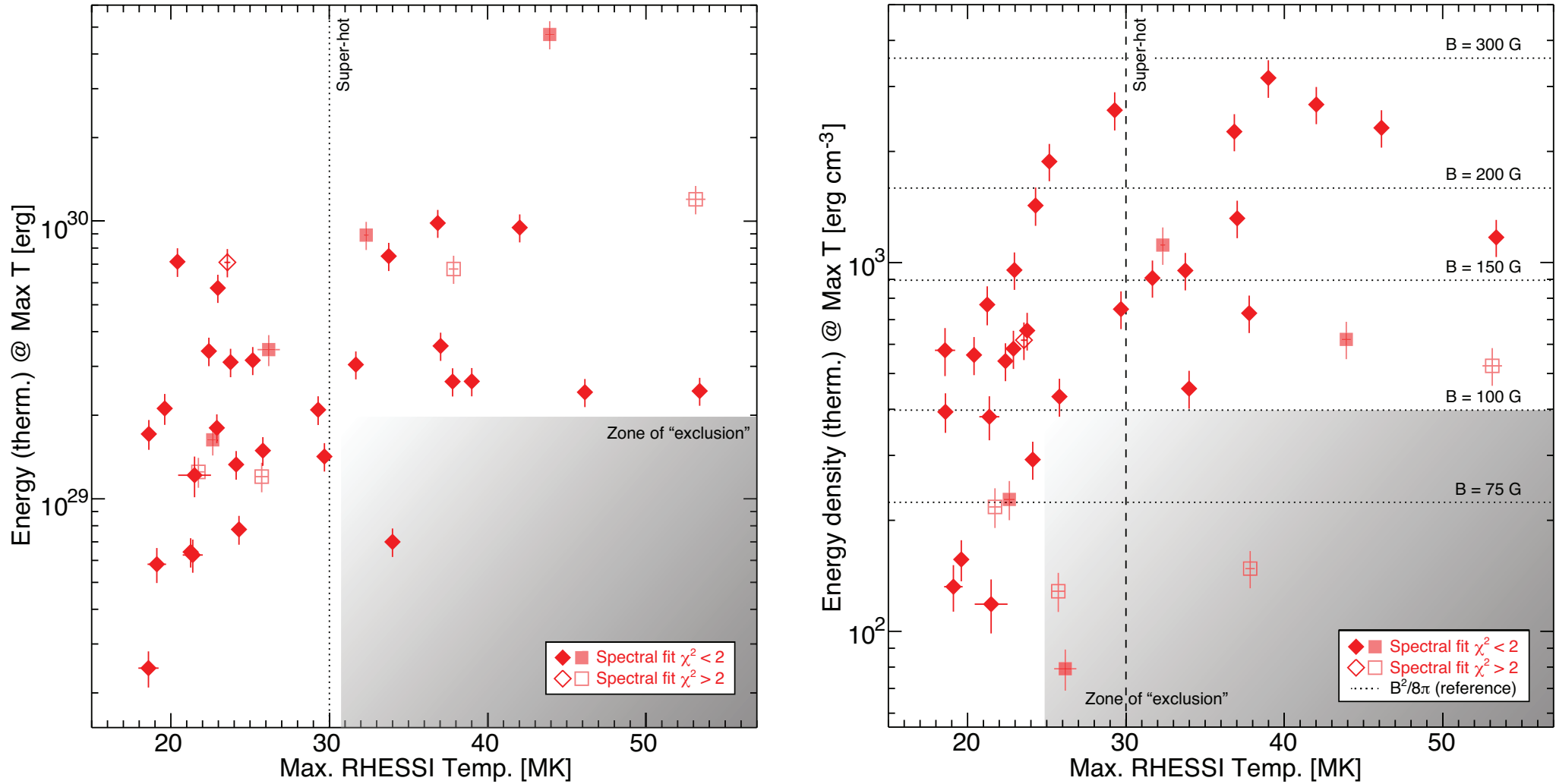

**Figure 4.** [*left*] Total thermal energy (assuming $T_i = T_e$) cotemporal with, and vs., the maximum measured continuum temperature. Thirteen of 14 super-hot flares exceed ~2.4×10$^{29}$ erg at the time of the maximum temperature, with smaller energies appearing to be "excluded," vs. a significant scatter among cooler flares. As for Fig. 3, *squares* represent cases where multiple sources may be skewing the volume/density measurements. [*right*] Thermal energy *density* cotemporal with, and vs., the maximum measured continuum temperature. Magnetic field strengths for selected values of equivalent magnetic energy density ($B^2/8\pi$) are shown for reference (*dotted* lines); these are the minimum field strengths required to contain the thermal plasma (i.e., $\beta < 1$). Thirteen of 14 super-hot flares require $B \gtrsim 100$ G in the corona, where the super-hot plasma is located; this is illustrated qualitatively by the "zone of exclusion" (*shaded*).

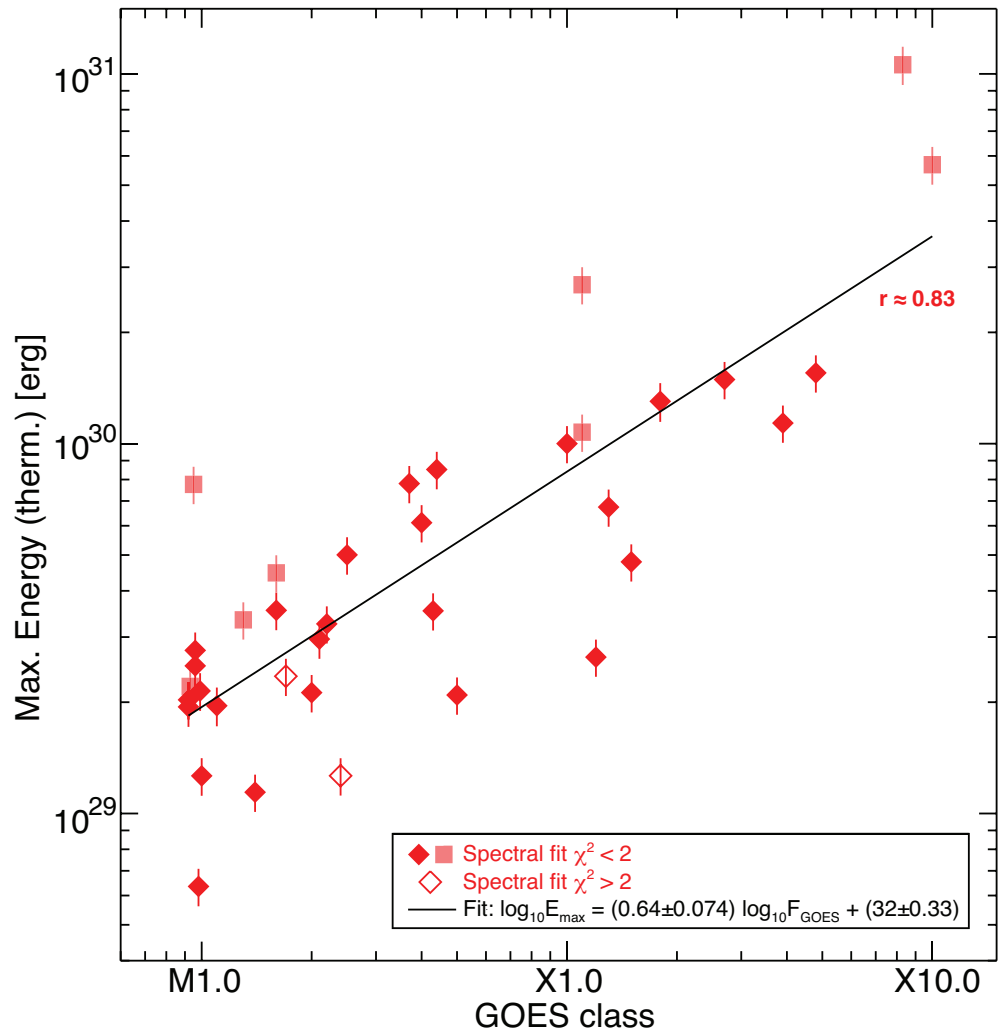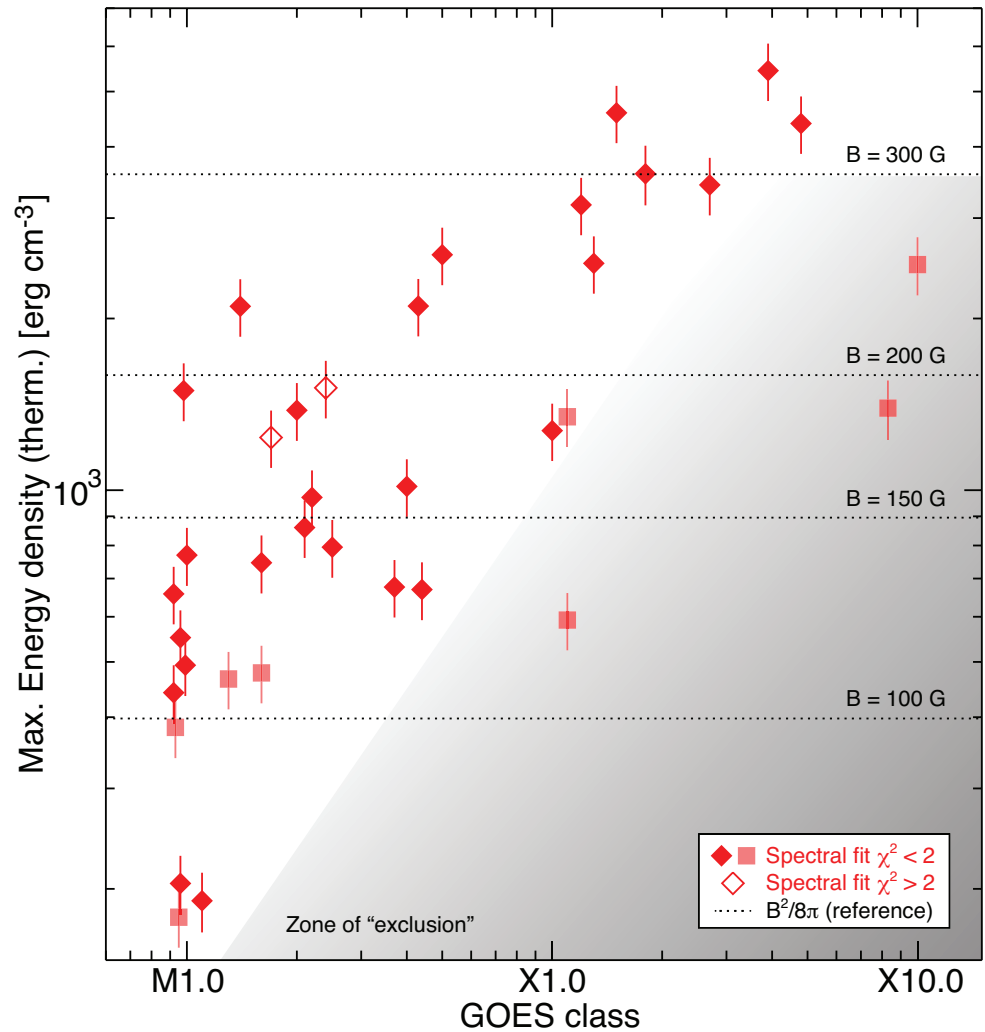

**Figure 5.** [*left*] Maximum total thermal energy (assuming $T_i = T_e$) achieved during the flare vs. GOES class. A strong power-law correlation is apparent. As in previous Figures, *squares* represent cases where multiple sources may be skewing the volume/density measurements. [*right*] Thermal energy *density* corresponding to the maximum energy vs. GOES class, with reference magnetic field strengths (*dotted* lines). Super-hot flares achieve significantly higher maximum energies and energy densities. Eleven of 12 X-class flares require $B \gtrsim 180$ G in the corona, as illustrated qualitatively by the "zone of exclusion" (*shaded*).

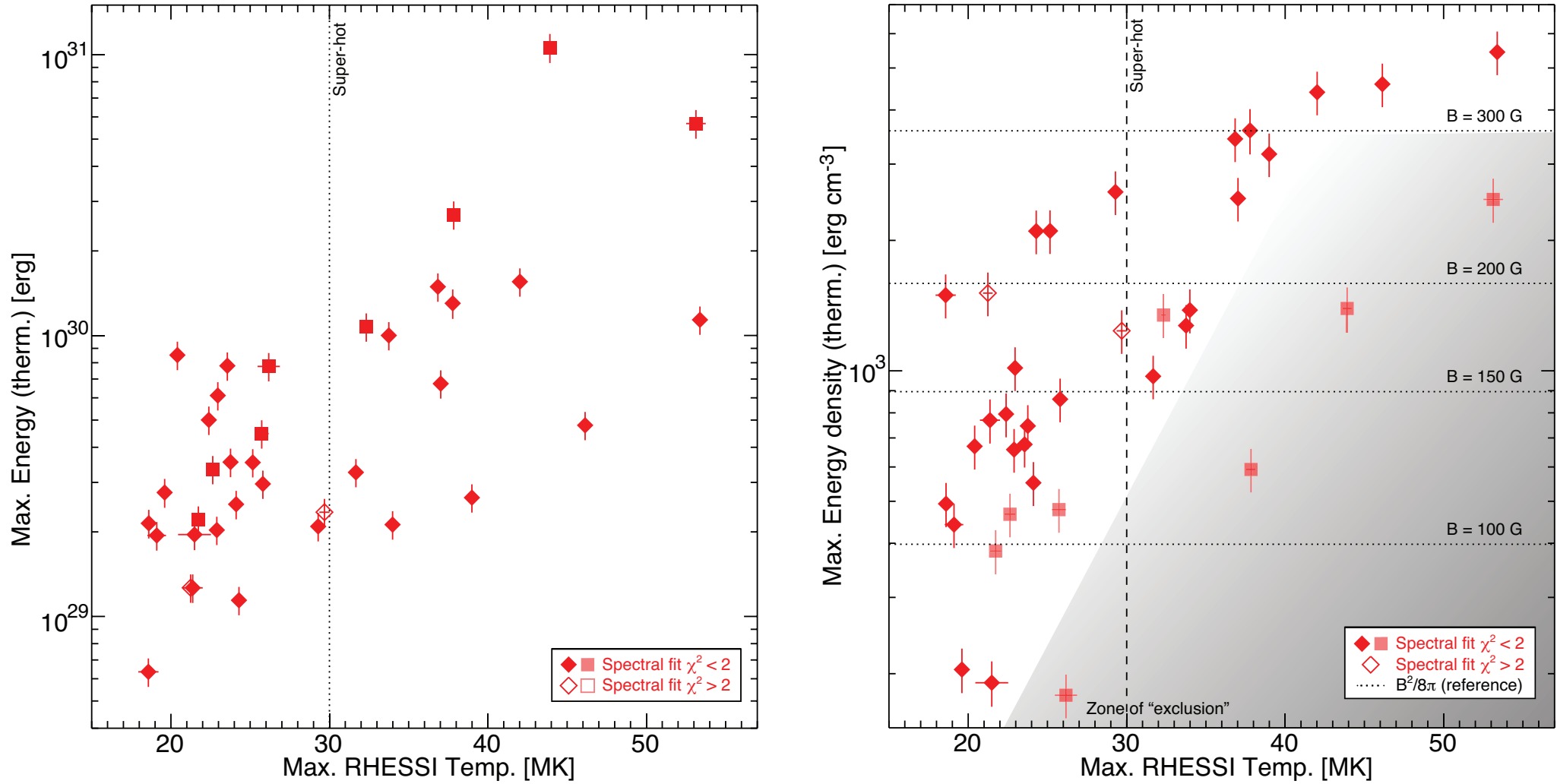

**Figure 6.** [*left*] Maximum total thermal energy (assuming $T_i = T_e$) achieved during the flare vs. maximum continuum temperature (*not* cotemporal). As in previous Figures, *squares* represent cases where multiple sources may be skewing the volume/density measurements. [*right*] Thermal energy *density* corresponding to the maximum energy versus maximum temperature (*not* cotemporal), with reference magnetic field strengths (*dotted* lines). *None* of the non-super-hot (<30 MK) flares exceeds ~$10^{30}$ erg, while 9 of 14 super-hot flares do. More strikingly, super-hot flares have significantly higher maximum energy density, with 13 of 14 exceeding ~970 erg cm$^{-3}$, equivalent to $B \gtrsim 160$ G; all super-hot outliers are from "possible multi-source" images (*squares*), and excluding these further strengthens the association, illustrated qualitatively by the "zone of exclusion" (*shaded*).